# REVIEW ON MASTER PATIENT INDEX

Dr W.G Prabath Jayatissa
Prof Vajira H W Dissanayake
Dr Roshan Hewapathirane

Post Graduate Institute of Medicine University of Colombo, Sri Lanka

*ABSTRACT*

*In today's health care establishments there is a great diversity of information systems. Each with different specificities and capacities, proprietary communication methods, and hardly allow scalability. This set of characteristics hinders the interoperability of all these systems, in the search for the good of the patient. It is vulgar that, when we look at all the databases of each of these information systems, we come across different registers that refer to the same person; records with insufficient data; records with erroneous data due to errors or misunderstandings when inserting patient data; and records with outdated data. These problems cause duplicity, incoherence, discontinuation and dispersion in patient data. With the intention of minimizing these problems that the concept of a Master Patient Index is necessary. A Master Patient Index proposes a centralized repository, which indexes all patient records of a given set of information systems. Which is composed of a set of demographic data sufficient to unambiguously identify a person and a list of identifiers that identify the various records that the patient has in the repositories of each information system. This solution allows for synchronization between all the actors, minimizing incoherence, out datedness, lack of data, and a decrease in duplicate registrations. The Master Patient Index is an asset to patients, the medical staff and health care providers.*

## 1. HISTORY OF MASTER PATIENT INDEX

History goes way back to 1950's when Automatic linkage of vital records by Newcombe (1) introduce how to bind the records through and get the unique combine records. There he tries to manage how to pair the surnames by using both father's surname and mother's maiden name. This article he suggested that the extent of efficiency of using surnames is more efficient than using the family group names to which two surnames are more efficient than one for identifying a family group has probably not been generally recognized.

Design and implementation of a data base for medical records done in 1980's by Larry & Fleming (2) of the College of Health University and the department of Computer Science of Florida in the United States of America. Larry and Colleen (1) suggested data base for the medical record first time in the history of the human kind. Those days computer systems were not developed as such today by their effort of trying to develop a health record system must be highly appreciated. The medical record has been identified as the source of data for a computerized medical information system. Larry & Colleen (1) suggested a design of a computerized database for structuring some of the data normally re-corded in the manual medical record. This design identifies objects of the health record which contain the computerized patient record. A method of transferring the computerized patient record to a low-cost method such as magnetic medium so calls old days large floppy disks and low volume hard disks at the time the patient is transferred from one health care institute to another healthcare institution. Low-cost hardware features were used to implementation of the above-computerized patient record for a health care institutes.

Dental Research: An International Journal (DRIJ)Vol.1, No.1

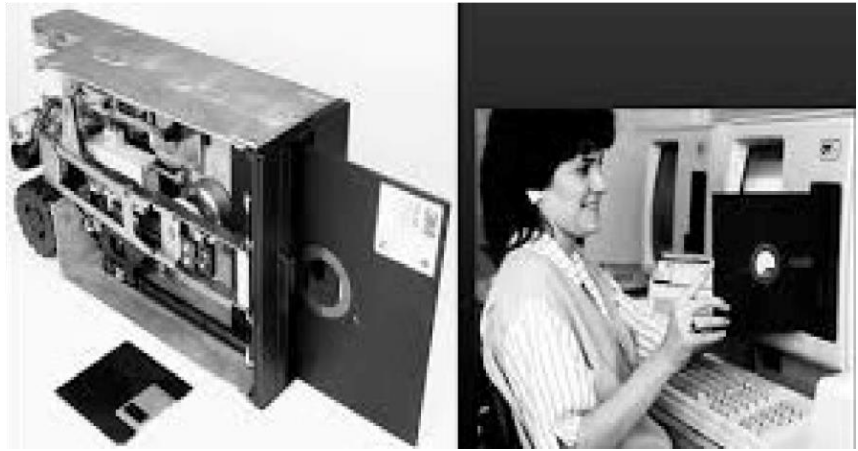

Figure 1 Use of Large Floppy Disk In 1980

In 1995 develop the National Patient Master Index (NPMI) in Singapore they have invented following things, Patients who are gone to the local hospital who has to take identification number will be detected so the duplication will have avoided (3). Federation of the personal identification services between enterprises by David W. Forslund (4) in year 2000 discuss about the managing the identities of persons within the particular organization or a domain (4). In 2012 Joffe E, Bearden CF (5), presented to the AIMA Annual Symposium about the implication for missed laboratory results following Duplicate patients in the system record. In August 2012 Sarvesh Jain (6) develop a prototype for Health Master Person Index for public use, he designed a detail architectural design to solve the problem with related to the patient data sharing in the different domain system. A high-level architecture of the Secure Open Enterprise Master Patient Index(SOEMPI) introduced by Toth and Durham (7) of University of Texas at Dallas introduce for the private patient record linkage. The latest implementation of master patient index in South Korea in 2015 introduce a Common Health Information Exchange (HIE) Platform with a national master patient index (8).

## 2. DUPLICATION OF PATIENTS IDENTITY

In 1995 develop the National Patient Master Index (NPMI) in Singapore they have invented following things, Patients who are gone to the local hospital who has to take identification number will be detected so the duplication will have avoided (3). For the security and the safeguard for the sensitive information of the patient smartcard, thumbprint retinal scan was introducing but some used places were unable to maintain them properly due to excess cost. After all these considerations Singapore government decided to use National Registration Identification Card (NRIC). NPMI was open to both government and the private sector. All the hospitals were allocated to the system and one senior administrative person is allocated for institutional responsibility. Inward or clinic data authority was managed by an authorized person. At the conclusion in Singapore use NPMI and drug allergy and medical alert date in large scale.

Federation of the personal identification services between enterprises by David W. Forslund (4) in the year 2000 discuss the managing the identities of persons within the particular organization or a domain. Duplication of records occurs when patient registered in multiple health care institutions or in hospitals. As a result of these multiple patient registrations, clinicians will come across with wrong information causing harmful prescriptions, procedures or therapies. There several studies related to reducing the duplication of patient registrations. So the legal issues has arrived this duplication of record in the past.

In 2012 Joffe E, Bearden C.F(5), presented to the AIMA Annual Symposium about the implication for missed laboratory results following Duplicate patients in the system records. In that same symposium discussed this case scenario, a case of two records for the same 17 years old patient, one record had several drug allergies and had admitted with anaphylactic shock. The second record is a completely a normal patient who is not an allergy to any known drug or food. During the merging of the true clinical history deleted and the false duplicate persisted in the system. This teenager admitted to a hospital after a fall from a skateboard with a lateral malleolus fracture. At the hospital diclofenac 100mg suppository administered per rectally. After several minutes' patient went into anaphylactic shock and treated at ICU for several days. After following this incident, the possible matches manually reviewed. However, manual review is a costly and time-consuming procedure.

In 2012 Aditya Pakalapati (9) of Utah University produce a content management facility(CMF) as the solution for MPI. In his dissertation, he suggested the that suggested CMF will restrict the data retrieval from the MPI only by the consent of the patient so the only the patient has the right to add, update or delete any data. So, with suggested CMF patient, privacy is protected. The only possibility that the MPI administrator allow is to remove the confirmed death patient record after the death record being entered into the system.

After review 20 000 randomly selected record-pairs from the set of potential duplicates algorithm was defined by Erel Joffe and Michael Byrne (5) in 2014. After doing the previous literature baseline set of parameters were identified. Base line parameters that were used are Name (first, middle, last) DOB (date of birth), SSN (social security number), gender (Male, Female), address and phone number. After the preliminary data review and algorithm identified and programmed.

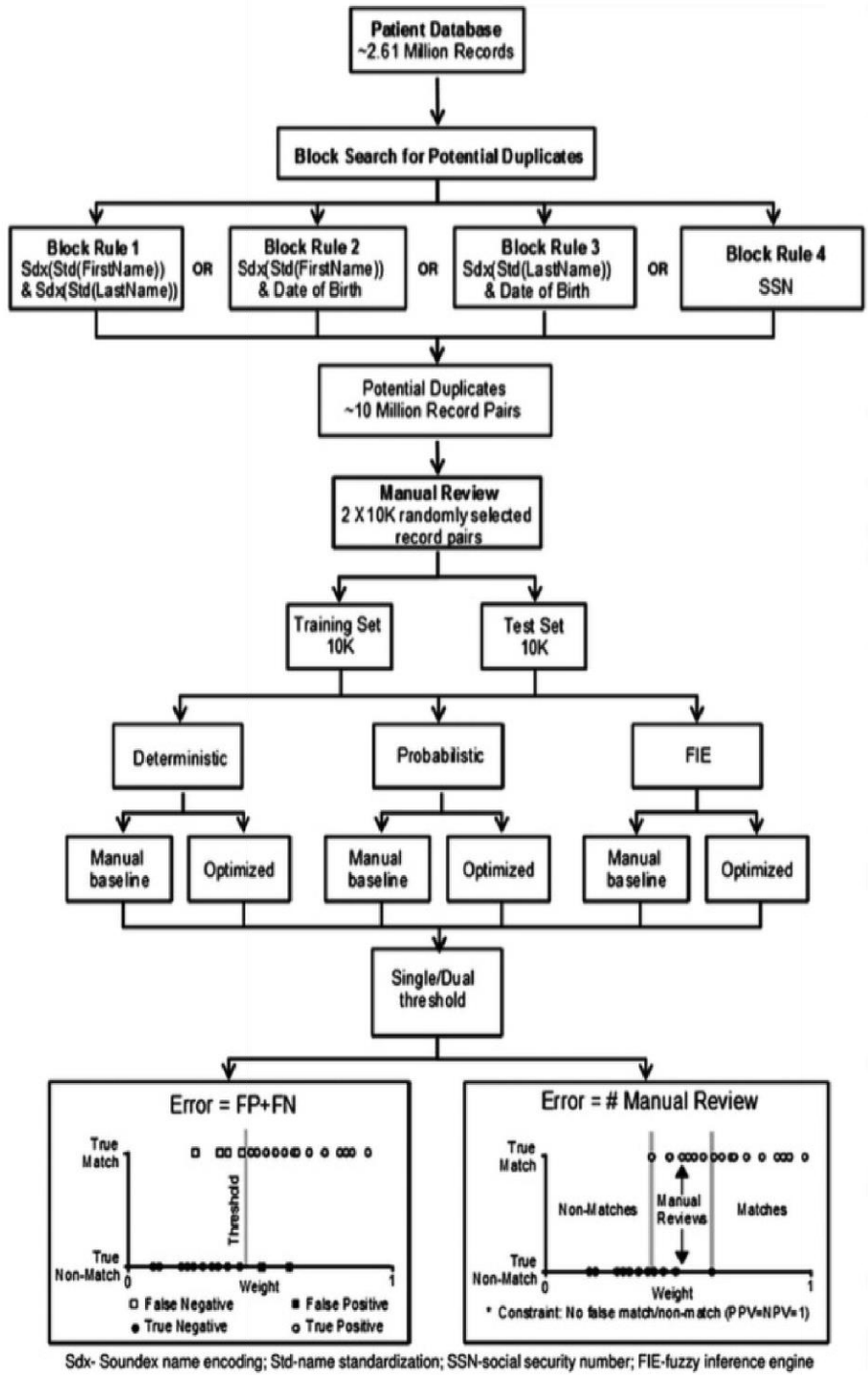

Figure 2 Algorithm Defined by Erel Joffe And Michael Byrne

## 3. MINIMUM CLINICAL DATASET (MCDS)

Minimum clinical dataset (MCDS) or the Minimum data set(MDS) is an important topic in healthcare information exchange platforms. There is no clear unified definition to MDS and this is poorly defined during the past few decades. The following definition is made for MDS. MDS is a coherent set of explicitly defined data elements in health care (23). MCDS is first introduced in the United State of America as the federally mandated process by the federal government for clinical assessment of all residents in health care. Commonly for Developing Minimum Clinical Datasets developer used consultants in different filled. While doing the discussion among these expert developers have to read through the literature reviews of former implementations and collect the existing clinical data set (8).

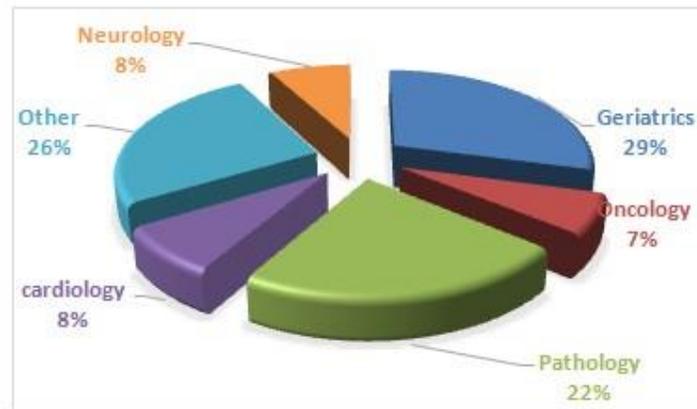

Figure 3 Publications by Minimum Dataset Clinical Specialty (8)

In August 2012 Sarvesh Jain (6) developed a prototype for Health Master Person Indexfor public use, he designed a detail architectural design to solve the problem with related to the patient data sharing in the different domain system. Integration infrastructure will provide the best software solution for interoperability of health information systems.

This integrated health information system improves health data sharing and provides better health services to the patient by connecting multiple isolated health domains to external systems. During the prototype, development developer tried to minimize the duplication and introduce a solution for a unique identification method. If there are several different first names of a pediatric patient, the merging application will merge simply all of them. This merging method called as stacking (11).

Some other MPI mergers determine the best accurate and mostly used scenarios. Also giving the chance for the manual record filtering using the expert on the filled. Intergraded systems much useful for the better healthcare but in some occasion, there was some adverse effect with these systems in 2004 Dr. J.T Finnell (12) published in his publication mentioned about the adverse effect following the mistreatment at the emergency departments.

The patient comes with feature suggestive of anterior lateral STEMI to the emergency department that needs thrombolization with thrombolytic. Treating patient doctor gone through the system and find the patient is treated for an intracranial hemorrhage(ICH) very recently. So, the doctor's conclusion was that with the thrombolytic treatment patient may start bleeding intracranial again.

He suggested not to thrombolysis and going for the interventional treatment. That treatment method is very expensive so the patient has to pay an extra amount of money in his old age. But after the intervention doctor gets to know that this patient was not treated for any ICH previously. This all due to the duplication of record in the system.

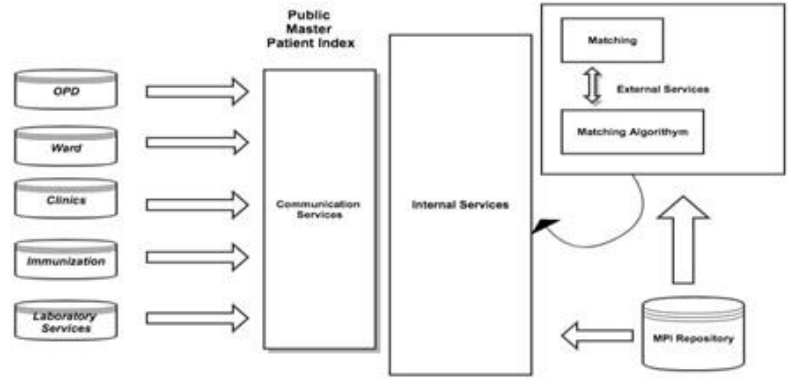

Figure 4  Development Methodology for Minimum Clinical Dataset Piper Svensson Ranallo And Terrence J. Adam

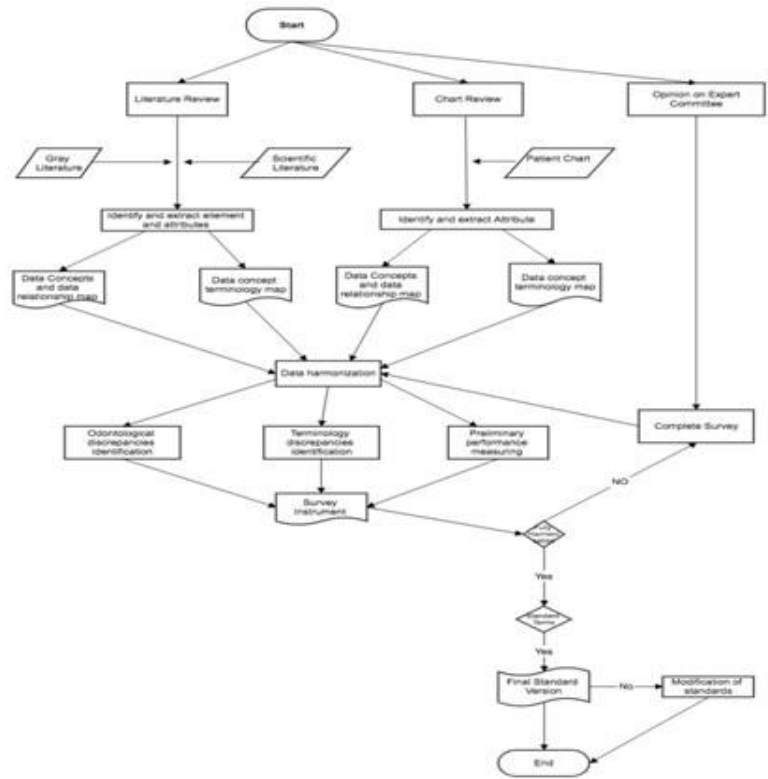

Figure 5 Development Methodologies for Minimum Clinical Dataset by Piper Svensson Ranallo And Terrence J. Adam.

## 4. MPI IMPLEMENTATIONS

In South Korea in 2015 introduce a Common Health Information Exchange (HIE) Platform with a national master patient index (8). Poor data quality is another major issue when the introduction of intergraded system. In one health care institute the data quality is in the standard format and others not. When comes to the integration, matching of health data in the different system will cause a major panic situation in the systems.

A high-level architecture of the Secure Open Enterprise Master Patient Index(SOEMPI) introduced by Toth and Durham (7) of University of Texas at Dallas introduce for the private patient record linkage. During the introduction of this high-level architecture emphasized the open source software frame work solve the most of the problem related to the private record linkage.

Aditya Pakalapati(9) of Utah State University implemented the content management system with a public health MPI(phMPI) for the Utah Department of Health. This phMPI can manage several entities of data by adding patients, updating them or deleting them. A unique identifier is also produced. Following versions graphical user interface enabled.

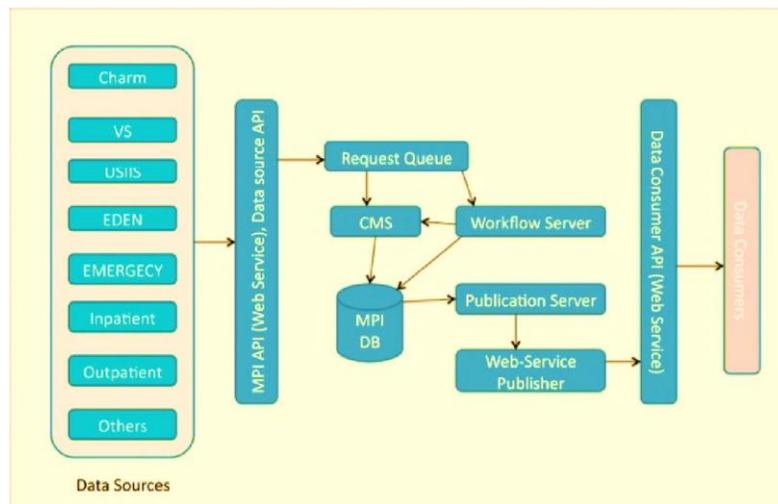

Figure 6  Overall Architecture of the PhMPI by Aditya Pakalapati Utah State University

## 5. UNIQUE PATIENT IDENTIFIER

In the US health care system identity crisis was identified by the Richard Hillestad, James H. Bigelow, Basit Chaudhry (13) and they suggested the benefits of a unique patient identifier for the U.S. Health Care System. Following Common mechanism of MPI introduction architecture introduce by the Pedro Tiago Magalhaes Gomes Master Patient Index Siemens group of hospitals in Portugal.

In 2014 patient identification and matching final report done by Genevieve Morris, Scott Afzal, Jan Greene, Chris Coughlin. Regarding the patient Privacy-Preserving of data integration in Health Surveillance research done by Jun Hu mentioned privacy preservation is more important as patient information security.

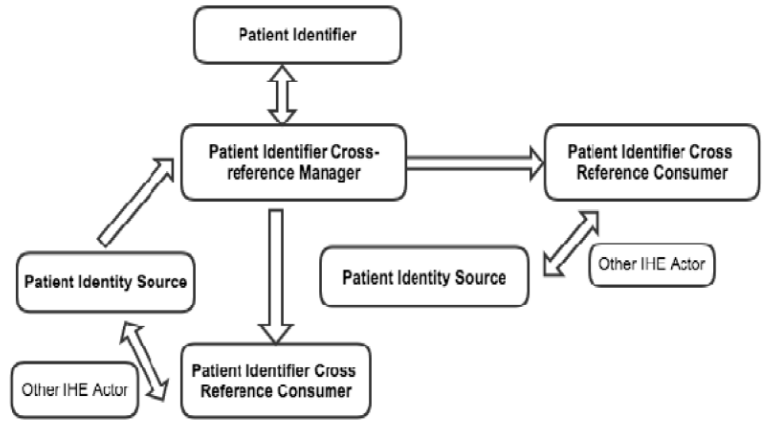

Figure 7 MPI by Pedro Tiago Magalhaes Gomes - MPI

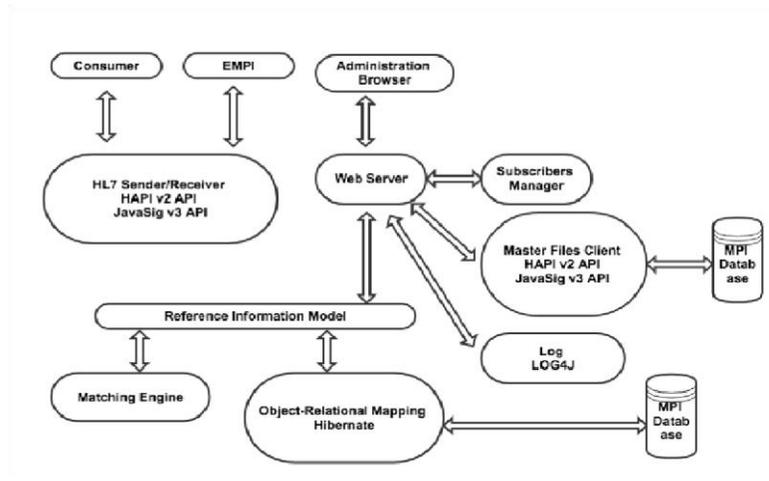

Figure 8 MPI by Pedro Tiago Magalhaes Gomes - Architectural Diagram

A unique identifier is also produced by Aditya Pakalapati of Utah State University implemented the content management system with a public heath MPI(phMPI) for the Utah department of Health(11). This phMPI can manage several entities of data by adding patients, updating them or deleting them. Following versions graphical user interface enabled.

## 6. DISCUSSION OF THE REVIEW

Indexing of the patient going back to 1950 while the using of the modern technology and making it to work in a small place like in a computer took 3 decades. Main concern of the MPI developers were the introduction of the unique identification methods using unique identifier and then interoperability among the health information systems using HL7 based or XML based. As a major solution for the duplication lot of counties have introduce a unique card where all the demographic data as well as the clinical data are stored. Duplication correction is a main concern in the evolving MPI systems in the world. While some developed computation method like algorithms to solve this common problem. Development of the minimal clinical dataset is a major part involving the medical professional where the most suitable clinical data set should be taken as the data set. But in a complete electronic health record the overlapping of these data sets are more common.

Getting all the systems of a health establishment to be synchronized in relation to their patients, will bring a greater value in the evaluation of a person's clinical condition, even if it is the patient's first contact with the health professional in. The MPI system in order to provide all these advantages has to implement the following basic functionalities:

- Ability to add people
- Ability to change existing data
- Ability to add identifiers to an existing person
- recognize equalities in records
- Join two registers in one so respond to surveys posted to the repository with a list of possible outcomes;
- synchronize associated systems
- respond to all actions in real time and without human help.

While aiming to mark an important step in the evolution of information systems. MPI cannot leave aside some important features

- be self-contained, so that it can easily be implemented in any health facility;
- be able to adapt to existing systems, not against the other;
- use all standards defined and accepted by the community in order to achieve
- communicate with all entities;
- comply with all your basic functions, since no system is successful if it is incomplete.